\documentclass[]{mn2e}
\usepackage{psfig}

\def\dm{$\delta$m~}
\def\deg{$^\circ$}

\begin{document} 

\title[Periodic variability in Kelu-1]{Periodic photometric variability of the
brown dwarf Kelu-1}

\author[F.J. Clarke, C.G. Tinney and K.R. Covey]{F.J. Clarke$^1$,
C.G. Tinney$^2$ and K.R. Covey$^3$\\ $^1$Institute of Astronomy,
Madingley Road, Cambridge CB3 0HA, UK.\\ $^2$Anglo-Australian
Observatory, Epping, Australia.\\ $^3$Carleton College, MN, USA.\\
email: fclarke@ast.cam.ac.uk, cgt@aaoepp.aao.gov.au}

\maketitle

\begin{abstract}
We have detected a strong periodicity of 1.80$\pm$0.05 hours in photometric
observations of the brown dwarf Kelu-1. The peak-to-peak amplitude of the
variation is $\sim$1.1\% (11.9$\pm$0.8\,mmag) in a 41nm wide filter centred
on 857nm and including the dust/temperature sensitive TiO \& CrH bands. We
have identified two plausible causes of variability: surface features
rotating into- and out-of-view and so modulating the light curve at the
rotation period; or, elliposidal variability caused by an orbiting
companion. In the first scenario, we combine the observed $v \sin i$ of
Kelu-1 and standard model radius to determine that the axis of rotation is
inclined at 65$\pm$12\deg\ to the line of sight.
\end{abstract}


\section{Introduction}

The study of rotation and variability in main sequence stars has led to a
great improvement in our understanding of their physics (e.g. Stauffer \&
Hartmann 1986). Recently, several groups have shown that variability can
also be detected in substellar brown dwarfs (Tinney \& Tolley 1999;
Bailer-Jones \& Mundt 1999, 2001; Mart\'\i n, Zapatero Osorio \& Lehto
2001).

In this paper we present differential photometry of the brown dwarf Kelu-1,
in a search for rotational variability. Kelu-1 is a field brown dwarf,
discovered by Ruiz, Leggett \& Allard (1997) via its large proper
motion. It is classified as an L2 dwarf in the scheme of Kirkpatrick et
al. (2000), and model fits to its spectrum estimate an effective
temperature of 1900$\pm$100K (Ruiz et al.\ 1997). The parallax distance of
19.6$\pm$0.5\,pc gives an absolute magnitude of M$_{\textrm{\small
j}}$=11.96$\pm$0.09 (Kirkpatrick et al.\ 2000) or a bolometric magnitude of
M$_{\textrm{\small bol}}$=13.9$\pm$0.1. The detection of strong lithium
absorption (Ruiz et al.\ 1997) in an object of this luminosity implies a
mass below 0.07M$_\odot$ (Tinney 1998), making Kelu-1 a bona fide brown
dwarf.  In common with many L- and M-type brown dwarfs (Basri et al.\ 2000,
Tinney \& Reid 1998), Kelu-1 is a very rapid rotator with a measured $v
\sin i$=60$\pm$5\,kms$^{-1}$. It is thought this may indicate that the
magnetic braking mechanisms which operate in more massive stars do not
operate with the same efficiency in brown dwarfs (as first suggested by
Tinney \& Reid 1998).

In section 2 we describe the data acquisition and reduction. Time
series analysis of the resulting differential photometry is presented
in section 3. In section 4 we discuss our observations in terms of
surface features on the brown dwarf, or the nature of an orbiting
companion.

\section{Data}

\subsection{Observations}

\begin{figure}
\caption{The spectrum of Kelu-1 (solid line) compared with the spectrum of
the L0 (slightly hotter) brown dwarf DENIS-P J0909-0658 (dashed line). The
transmission profile of the R6 filter is overplotted (dot-dash line),
showing how it isolates the sensitive TiO and CrH features at
$\sim$850nm.}
\label{fig:specandfilter}
\psfig{file=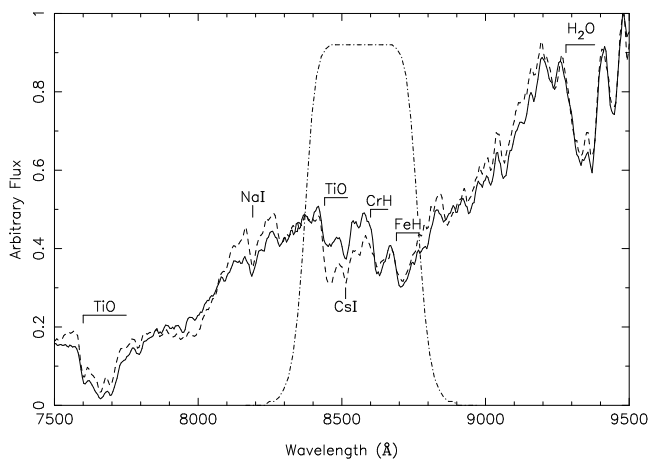,angle=270,width=8.5cm}
\end{figure}

Kelu-1 was observed on 2000 March 24-25 using the Taurus-2 instrument on
the 3.9m Anglo-Australian Telescope with a CCD detector denoted
MITLL3. MITLL3 is a 2096$\times$4096 pixel device with 15$\mu$m pixels,
giving an image scale of 0.373" per pixel. Observations were performed
through an intermediate bandpass blocking filter for the Taurus Tunable
Filter.  This filter (denoted R6) is 41nm wide and is centred at
858nm. The bandpass of the filter is shown overlaid on a spectrum of
Kelu-1 (taken from Mart\'\i n et al.\ 1999) in
Figure~\ref{fig:specandfilter}. Also plotted in
Figure~\ref{fig:specandfilter}~is the spectrum of the slightly hotter L0
dwarf DENIS-P J0909-0658 (Mart\'\i n et al.\ 1999), indicating the
temperature sensitivity of the TiO and CrH molecular bandheads selected by
our filter.

\begin{figure}
\caption{``Slot'' mask image of Kelu-1 and the {\em comparison} and {\em
check} stars used in our differential photometric analysis. The vertical
bars in the image are bad columns on the MITLL3 detector. The orientation
of the sky through the mask when Taurus-2 is rotated to PA=140\deg\ is
indicated.}
\label{fig:kelu1finder}
\psfig{file=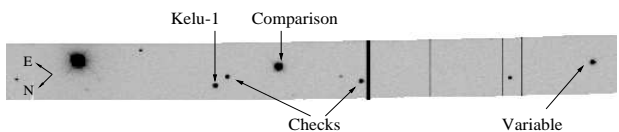,width=8.5cm}
\end{figure}

A specially constructed slot mask was installed in the focal plane wheel of
Taurus-2 (the same mask used by Tinney \& Tolley 1999). Multiple exposures
of a sky field can then be made through this slot, between which charge is
shuffled and stored on the unilluminated surface of the CCD. When used to
observe Kelu-1, the entire Taurus-2 instrument was rotated at an angle of
140\deg, so that several suitable photometric reference stars could be
observed (see Figure~\ref{fig:kelu1finder}). This observing technique
results in a series of 30 exposures, following which the CCD is read out
only once, providing very low read-out overheads and precise
timing. Table~\ref{tab:observationlog}~gives a log of the observations.

\begin{table}
\caption{Log of Kelu-1 observations in March 2000.}
\label{tab:observationlog}
\begin{tabular}{ccccc}
Date&Start UT&End UT&\# Images&Exposure Time (s)\\
\hline
24$^{th}$&11:20&11:53&60&60\\
24$^{th}$&13:15&14:00&90&30\\
24$^{th}$&15:04&19:05&480&30\\
25$^{th}$&10:10&12:56&270&30\\
25$^{th}$&13:39&19:00&300&60\\
\end{tabular}
\end{table}

\subsection{Reduction and photometry}
\label{sec:ReductionAndPhotometry}

Each CCD frame contains 30 seperate exposures, so the data processing is
slightly more complicated than for standard CCD imaging. Each {\em frame}
was bias subtracted and then sliced into thirty individual {\em
exposures}. The header file for each {\em exposure} was modified so as to
include the correct timing and positional information. A master flatfield
{\em exposure} was then constructed from dome flats observed in the same
manner as the data. The data were then divided by this flat, providing
correction to the $\sim$1\% level.

Due to the proximity of a nearby star to our primary target, point-spread
function (PSF) photometry of this data is required. Photometry was
performed using DAOPHOT within the IRAF environment. A PSF was determined
for each individual exposure by fitting a profile to several of the
brightest (non-saturated) stars in the field. This profile was then scaled
to all the stars in the field to determine their brightness. No faint
neighbours were found after the PSF had been subtracted from the target
stars. Pixels which varied by more than 5\% from the mean in the flatfield
were flagged as bad. These bad pixels were interpolated over during the PSF
fitting stage, but completely ignored during the actual photometry. PSF
photometry also has the advantage of being relatively immune to cosmic ray
events, which would affect aperture photometry. In order to verify our PSF
photometry, we also performed aperture photometry in the same fashion as
Mart\'\i n et al.\ (2001). The aperture photometry was found to give the
same, but noisier, results as the PSF photometry.

To obtain high precision differential photometry of Kelu-1, we compared its
apparent brightness with that of surrounding reference stars, cancelling
the effects of apparent brightness changes due to variable extinction,
seeing, instrument performance or exposure time. There are four available
reference stars in our field-of-view (the brightest star in the field was
often saturated).  We define the brightest of these four stars to be our
{\em comparison} star, which sets the magnitude zero-point for each
exposure. The three remaining stars can then be used as {\em check} stars,
allowing us to detect intrinsic variability in our {\em comparison}
star. One of the {\em check} stars was found to be variable and therefore
discarded. The remaining {\em checks} were averaged to define a mean {\em
check} magnitude. These stars are identified in Figure~\ref{fig:kelu1finder}.

\begin{figure}
\centering
\caption{The top panel shows differential photometry for Kelu-1 (in the
sense Kelu-1 minus {\em comparison} as described in \S
\ref{sec:ReductionAndPhotometry}) over two nights. The bottom panel shows
{\em comparison} minus {\em check} to the same scale, clearly indicating
the variability detected is from Kelu-1, not the {\em comparison}. The
uncertainties plotted are based on the photon counting errors as produced
by DAOPHOT and propagated to the differential results for Kelu-1, {\em
comparison} and {\em check} stars. This plot shows raw differential
photometry, before the correlation of Figure~\ref{fig:ycorrelation} is
removed.}
\label{fig:lc}
\psfig{file=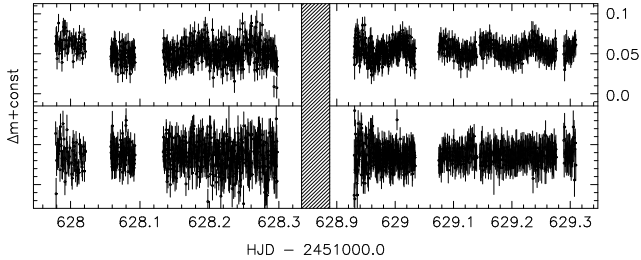,angle=270,width=8.4cm}
\end{figure}

The differential lightcurves obtained are shown in Figure~\ref{fig:lc}. The
top panel shows Kelu-1 minus {\em comparison}, and the lower panel {\em
comparison} minus {\em check}. It is clear from these plots that Kelu-1
minus {\em comparison} photometry is more variable than {\em comparison}
minus {\em check} photometry - indicating the variability detected is due
to changes in the apparent brightness of Kelu-1 itself (an increase in \dm
represents a dimming of Kelu-1). Variability in the {\em comparison} star
would manifest itself as equal, but opposite, changes in each
lightcurve. This effect can be seen on the first group of datapoints - the
Kelu-1 minus {\em comparison} data is above the mean, and the {\em
comparison} minus {\em check} photometry is below the mean, indicating the
{\em comparison} star has dimmed. We neglect this group of (60) datapoints
in further analysis.

As the photometry presented in Figure~\ref{fig:lc} is differential, most
sources of systematic error which might produce a ``spurious'' signal in
Kelu-1 are cancelled out. The remaining possible sources of error are
differential effects due to the differences in colour of Kelu-1 and the
reference stars, or their different locations on the CCD.

\begin{figure}
\caption{Differential photometry plotted against Y pixel position on the
CCD. There is a clear correlation at the level of 0.003mag per pixel, which
we remove by subtracting a straight line fit.}
\label{fig:ycorrelation}
\psfig{file=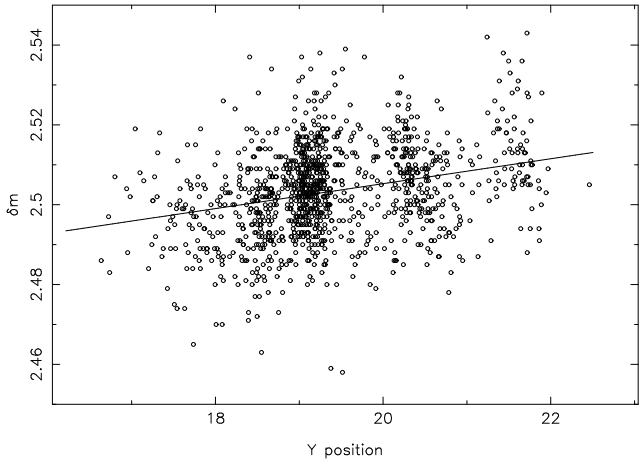,angle=270,width=8.4cm}
\end{figure}

Differential position effects have been minimized by keeping all objects
within 30 pixels of a nominal position on the detector for all
observations. However, differential photometry for Kelu-1 versus detector
position shows a residual correlation between \dm and y-position on the CCD
(Figure~\ref{fig:ycorrelation}). We have fitted this correlation with a
straight line and subtracted it. The cause of this effect remains
unknown. The maximum pixel-to-pixel sensitivity variation in the flat field
is 2.5\% in the region of interest, and does not display any gradient. All
data discussed hence have had this correlation removed. No correlation was
found with x-position.

\begin{figure}
\caption{Differential photometry plotted against airmass for Kelu-1 minus
{\em comparison} (top panel) and {\em comparison} minus {\em check} (bottom
panel). A Rank-Spearman correlation test shows no significant correlation
between airmass and differential magnitude.}
\label{fig:photvsam}
\psfig{file=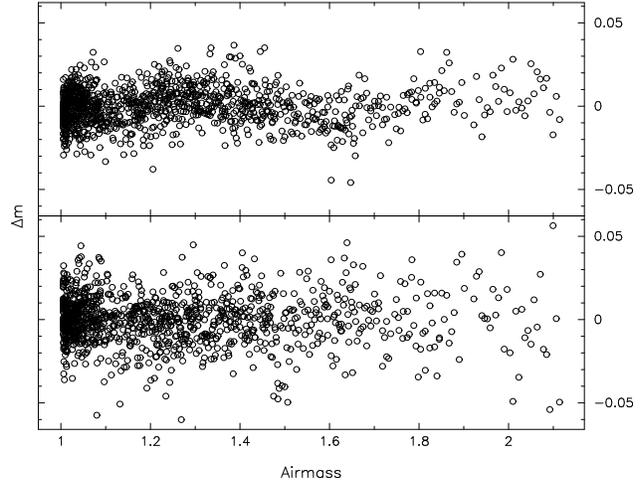,angle=270,width=8.4cm}
\end{figure}

Differential colour effects will be produced when the effective wavelength
of a star through a filter is different from the reference objects. The
target will then suffer a different amount of atmospheric extinction as the
target rises and sets. If such an effect is present we would expect that
spurious periodicities would be produced at aliases of 24 hours (ie 6,
12 or 24 hours). We would also expect to observe a correlation between
differential photometry and airmass if this effect is
present. Figure~\ref{fig:photvsam} shows a plot of the differential
photometry for Kelu-1 (upper panel) and comparison stars (lower panel)
versus airmass. There is no evidence (from Rank-Spearman tests) for a
correlation in either of these plots. In addition,
Figure~\ref{fig:specandfilter} shows that Kelu-1 does not have a strong
colour gradient over our filter band-pass.

\section{Time Series Analysis}

We have calculated the least-squares weighted power spectrum of the
differential lightcurve to look for periodicities. The power spectrum is
calculated in the manner described by Frandsen et al.\ (1995), which we
briefly describe here. The data are represented by the function
$A_i\sin(2\pi f_it + \phi_i)$ for a range of frequencies $f_i$, where
$A_i$, $\phi_i$ and $t$ are the amplitude, phase and time respectively. The
power at each frequency is then $P_i = A_i^2 = \alpha^2 + \beta^2$, where
$\alpha$ and $\beta$ are given by

\begin{eqnarray*}
\alpha&=&(sc_2 - cx)/(s_2c_2 - x^2),\\
\beta&=&(cs_2 - sx)/(s_2c_2 - x^2)
\end{eqnarray*}

\noindent
with

\begin{eqnarray*}
c & = & \sum w_jx_j\cos(\Omega_it_j),\\
s & = & \sum w_jx_j\sin(\Omega_it_j),\\
c_2 & = & \sum w_j\cos^2(\Omega_it_j),\\
s_2 & = & \sum w_j\sin^2(\Omega_it_j),\\
x & = & \sum w_j\sin(\Omega_it_j)\cos(\Omega_it_j)
\end{eqnarray*}

\noindent
where ($t_j$,$x_j$,$w_j$) are the data points \& associated weights and
$\Omega_i=2\pi f_i$. Each data point is assigned a weight of $w_j=1/e_j^2$,
where $e_j$ is the error associated with point $j$. This system gives data
points with smaller errors higher weights.

The discrete sampling of observations means that the observed power
spectrum is the convolution of the ``true'' power spectrum, with the
``window function''. We have calculated the window function as prescribed
by Roberts et al.\ (1987);

\begin{equation}
W(f) = \frac{1}{N}\sum_j e^{-2\pi ift_j}
\end{equation}

\begin{figure}
\caption{The lower panel shows the least-squares weighted power spectrum of
Kelu-1 minus {\em comparison} (solid line) and {\em comparison} minus {\em
check} (dashed line). There is a clear peak at 13.33 day$^{-1}$
(period=1.80$\pm$0.05 hours) with a peak-to-peak amplitude of 11.9 mmag in
the Kelu-1 power spectrum. The error bar represents the 1-$\sigma$ width of
the central peak. There is no corresponding peak in the {\em comparison}
minus {\em check} power spectrum, showing that the detected period is from
Kelu-1. The middle panel shows the window function of the dataset, clearly
indicating the peak is not due to sampling effects. The upper-most panel
shows the least-squares weighted power spectrum of Kelu-1 over the entire
range of frequencies we can study, from 1 day$^{-1}$ to the Nyquist
frequency of 0.5 min$^{-1}$. This panel shows there is no significant
structure in the power spectrum above f$\sim$20\,day$^{-1}$}
\label{fig:powerspec}
\psfig{file=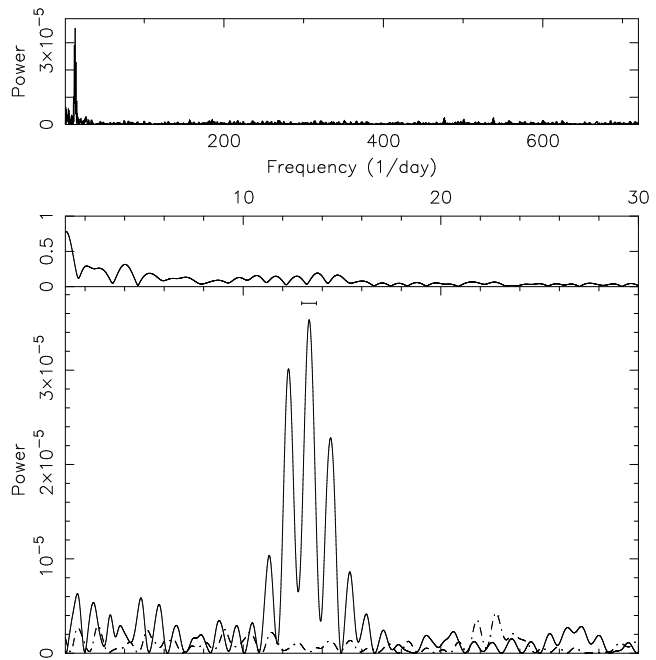,angle=270,width=8.5cm}
\end{figure}

The power spectrum and window function of our data are shown in
Figure~\ref{fig:powerspec}. The upper panel shows the power spectrum over
the complete range of frequencies available to us; one day$^{-1}$ up to the
Nyquist frequency of 0.5 min$^{-1}$ ($1 < f < 720$ day$^{-1}$; $24 > P > 0.033$
hours). There is only one main peak in the power spectrum. The lower panel
shows a close-up of the region of interest ($1 < f < 30$ day$^{-1}$). The
power spectrum of the Kelu-1 differential photometry (solid line) shows a
strong peak at a frequency of 13.34$\pm$0.38\,day$^{-1}$, corresponding to a
period of 1.80$\pm$0.05\,hours. The peaks at higher and lower frequencies
are produced by the first sidelobes of the window function, as shown in the
middle panel of Figure~\ref{fig:powerspec}. The maximum power of the main
peak is 3.5x10$^{-5}$\,mag$^2$, corresponding to a peak-peak amplitude in
the lightcurve of 11.9\,mmag. Monte-Carlo simulations of the data give a
1-$\sigma$ error in the amplitude of 11.9$\pm$0.8\,mmag. CLEANed power
spectra (Roberts et al.\ 1987) do not reveal any other periodicities hiding
in the sidelobes of the main frequency.

Also shown in the lower panel of Figure~\ref{fig:powerspec} is the power
spectrum of the {\em comparison} minus {\em check} photometry (dashed
line). There is no peak at the frequency detected in the Kelu-1 photometry,
and the maximum power in this spectrum is an order of magnitude lower. This
confirms the periodic variability we detect in the Kelu-1 minus {\em
comparison} photometry is due to intrinsic variability of Kelu-1 rather
than the {\em comparison} star.

\begin{figure}
\caption{Power spectrum of data from night 1 (upper panel) \& 2 (lower
panel). The same peak at $\sim$13 day$^{-1}$ is detected in both
datasets. The upper section in both panels shows the window function.}
\label{fig:nightlypowerspec}
\psfig{file=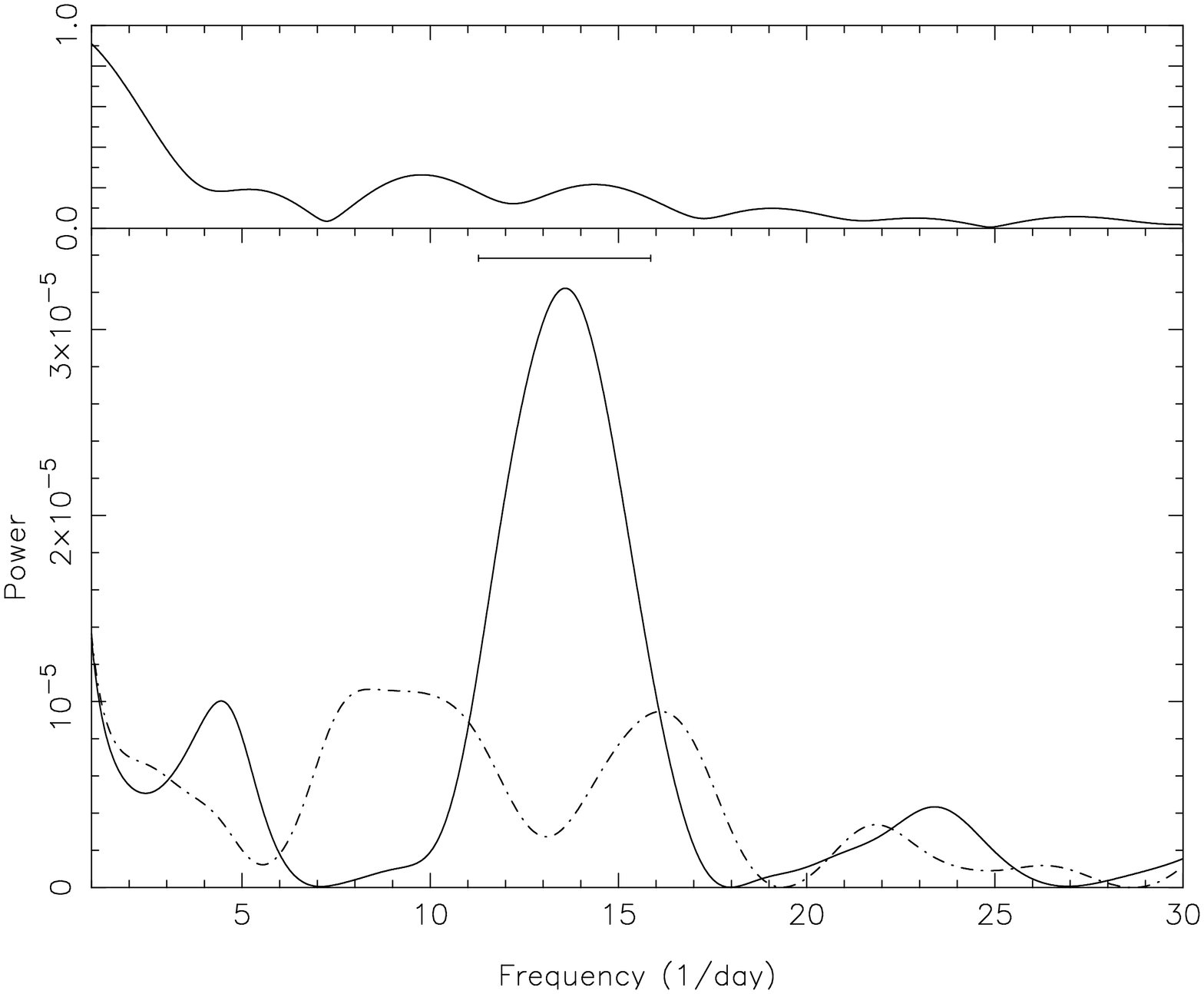,angle=270,width=8.4cm}
\psfig{file=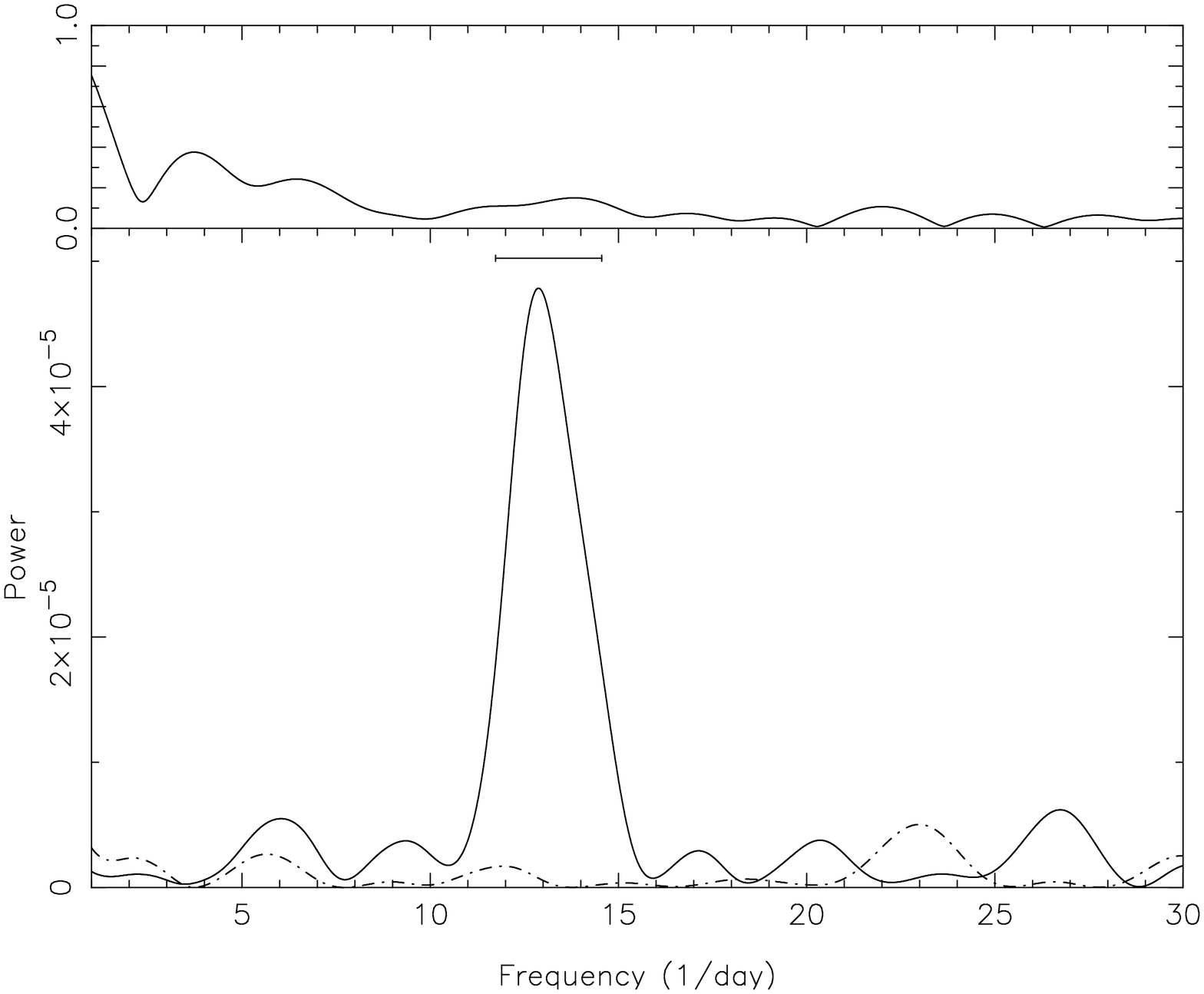,angle=270,width=8.4cm}
\end{figure}

To further investigate the periodicity detected, we have treated each
night's data separately. The periodograms are shown in
Figure~\ref{fig:nightlypowerspec}. The same $\sim$1.8 hour periodicity is
detected in both datasets, indicating the source of the variability is
stable over at least $\sim$48 hours ($\sim$25 rotations).

\begin{figure}
\caption{Kelu-1 photometry folded onto the main period (1.80 hours), and
combined into 5 minute bins to improve the signal to noise ratio. This data
has had a long term trend of \dm brightening at $\sim$3\,mmag/day removed.}
\psfig{file=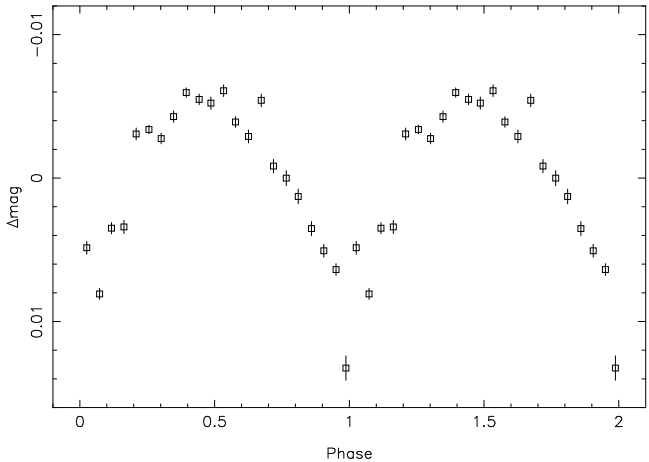,angle=270,width=8.5cm}
\label{fig:foldedlc}
\end{figure}

Close examination of the data reveals a slow linear change in Kelu-1's
brightness at a level of 0.003 mag/day over the duration of the
observations (this is not caused by the previously discussed correlation
between \dm and y-position on the CCD (Figure~\ref{fig:ycorrelation}). We
have {\em not} removed this effect for the time-series
analysis. Figure~\ref{fig:foldedlc} shows the Kelu-1 photometry folded on a
period of 1.80 hours (with the slow, linear trend this time removed), and
combined into 5 minute bins. We derive the following ephemeris for the time
of minimum light, t$_{\textrm{\small min}}$;

\begin{center}
t$_{\textrm{\small min}}$ = JD 2451628.037 + 0.075 $\times$ $\epsilon$
\end{center}

\noindent
where $\epsilon$ is the number of rotations since JD=2451628.037.

\section{Discussion}

Kelu-1 clearly displays periodic variability, but what is the cause of this
variability? We have developed four possible explanations which we discuss
in this section:

\begin{enumerate}
\item surface inhomegeneity moderated by meteorology and variable dust
formation;
\item surface inhomegeneity moderated by magnetic starspots;
\item light curve variability due to gravitational (i.e. tidal) distortion
of Kelu-1's envelope by a close companion; and
\item light curve variability due to an eclipsing binary.
\end{enumerate}

\subsection{Dust cloud meteorology}

At the cool temperature of Kelu-1's photosphere, dust formation will play
an important role. Theoretical spectra indicate that Kelu-1 is better
matched by models with dust suspended in the photosphere than models in
which condensates are ``rained-out'' (Baraffe et al.\ 1998). We cannot
expect, however, the atmospheres of brown dwarfs to be placid, homogenous
places --- especially if rotating at $v \sin i = $60\,km\,s$^{-1}$. They
are no doubt dynamic and evolving, much like the atmospheres seen on
planets in our own solar system. This leads us to investigate the
possibility that we are observing dust cloud meteorology in Kelu-1's
atmosphere.

Gelino et al.\ (2001) suggest there is a possible correlation between
J-K$_s$ colour and variability in L dwarfs -- variable objects tend to have
bluer J-K$_s$ colours, which they claim is evidence that {\em holes} in
dust clouds are causing variability. Kelu-1 has a J-K$_s \sim 1.6$
(Kirkpatrick et al.\ 1999). This is not significantly different from the
mean colour for an L2 dwarf (J-K$_s \sim 1.5$), so Kelu-1's strong
variability sheds no light on this suggestion.

Could the variability be due to a large feature on one side of the brown
dwarf? The surfaces of all the Solar System giant planets are dominated by
surface banding, which is driven by an interior in which Coriolis forces
dominate over buoyant convection, to produce long, thin quasi-cylindrical
cells oriented with their axes parallel to the axis of rotation (Schubert
\& Zhang 2000). In brown dwarfs, however, -- even brown dwarfs with a 1.8h
rotation period -- buoyancy forces should dominate to produce chaotic three
dimensional convection.  So banding structures, like those seen on Jupiter,
are probably not to be expected. If cloud formation is inhomogenous,
rotation and convection should combine to produce less ordered structures.
So convective upwelling, or large cloud inhomogeneities may be possible.

\subsection{Magnetic starspots}

Another possibility is that the imhomogeneities we detect are caused by
magnetically induced starspots, analogous to those observed on the Sun and
other cool stars. However, Gelino et al.\ (2001) have shown that the
atmospheres of brown dwarfs are unable to sustain starspots, as low
Reynolds numbers imply the atmosphere is not bound to magnetic field
lines. In addition, several groups have shown that there no strong evidence
for the link between variability and activity (as measured by
H-$\alpha$~equivelent width), which would be expected if magnetic forces
drove variability. We note however that this conclusion is weakened by the
small number statistics currently involved.

\subsection{Binarity}

One intriguing possibility is that we are seeing variability induced by a
companion orbiting Kelu-1. A companion could modify Kelu-1s brightness in
two ways; either by partial or total eclipsing, or via the gravitational
pertubation of Kelu-1's photosphere, causing ellipsoidal variability (see
Hilditch 2001). We have not been able to derive any orbital configuration
which could reproduce the observed lightcurve amplitude and shape, and we
therefore reject an eclipsing companion.
In the case of ellipsodial variability, we would observed two maxima per
rotation, and hence the orbital period would be 3.6 hours. It is reasonable
to assume the orbits will be circular due to tidal effects, and we can
therefore directly calculate the orbital seperation as;

\begin{equation}
a =
0.48R_\odot\left(\frac{M_1}{0.065M_\odot}\right)^{1/3}\left(\frac{P}{3.6\textrm{h}}\right)^{2/3}
\label{eq:orbitalsep}
\end{equation}

\noindent
where $M_1$ is the primary (Kelu-1) mass, and $P$ is the orbital period.

\begin{figure}
\caption{Allowed masses for a companion to Kelu-1. If Kelu-1 has a mass
greater than $\sim$45M$_{\textrm{\tiny jup}}$, the minimum mass limit is
set by the need to induce 1.1\% elliposidal variations with
i$\leq$65\deg. However, if Kelu-1 has a mass less than
$\sim$45M$_{\textrm{\tiny jup}}$, brown dwarfs/gas-giant planets are
excluded from the hatched region as they would overflow their Roche-lobe.}
\label{fig:secondarymass}
\psfig{file=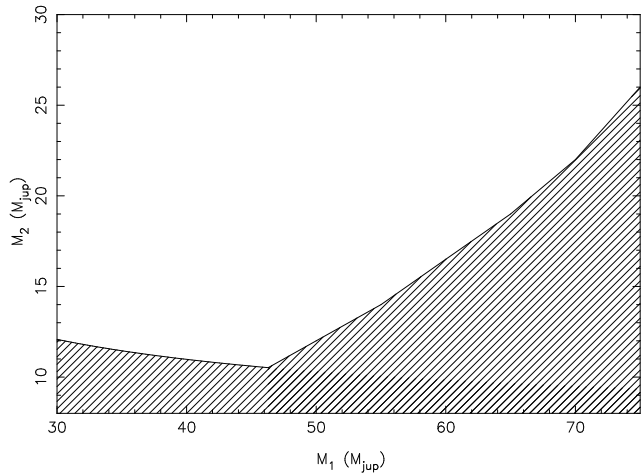,angle=270,width=8.4cm}
\end{figure}

Unfortunately, when fitting for ellipsodial variability, there is a
degeneracy between the mass of the secondary object and the orbital
inclination of the system. If, however, we assume the secondary companion
is also a brown dwarf (lack of X-ray flux rejects the possibility of a
massive dark companion such as a neutron star or black-hole), we can place
limits on its mass. Figure~\ref{fig:secondarymass} shows the possible
secondary mass as a function of primary mass. Companions are excluded from
the hatched region for two different reasons: (1) if
$M_1\geq45M_{\textrm{\small jup}}$, the minimum mass is determined by the
need to produce 1.1\% ellipsodial variability without the secondary causing
eclipses (i$\leq$65\deg), (2) for $M_1\leq45M_{\textrm{\small jup}}$, brown
dwarfs cannot exist in the hatched region, as they would overflow their
Roche-lobes, transfering matter to the primary. It is not clear what the
final outcome of such an interaction would be, but the lack of activity
indicates this cannot currently be happening. An object more compact than a
brown dwarf or gas-giant planet could survive in this region. We note that
these limits are based on simplistic models and are very sensitive to the
exact amplitude of Kelu-1's variability. They do however indicate that any
companion to Kelu-1 would be a brown dwarf rather than massive planet
(i.e. $M_2>13M_{\textrm{\small jup}}$).

Pure ellipsodial variability produces a sinusoidal lightcurve, which is
statistically consistent with Kelu-1's folded lightcurve. If a secondary
companion is responsible for the variability, future observations will
return the same amplitude and ephemeris. We note that a binary system would
have an orbital velocity of $\sim$130kms$^{-1}$, and would be capable of
mimicing the observed 60kms$^{-1}$ line broadening over a 1 hour
integration.

\subsection{Surface structures}

Doppler imaging codes have been used extremely successful to mapping and
understand the surface features on solar-type low-mass stars. However, the
process requires very high quality photometry and spectroscopy, combined
with a full understanding of photospheric physics; all of which seem a long
way off for brown dwarfs.

Although mapping the surface of a brown dwarf is a distant goal, we
can make some headway.  By synthesising photometry from theoretical
spectra (Allard et al.\ 2001) of two limiting cases; 1) dust is
suspended in the atmosphere ({\em DUSTY}), and 2) dust forms, but
immediately settles below the photosphere ({\em COND}), we can
investigate what degree of surface structure is required to reproduce
the observed variability.

Kelu-1's spectrum is best matched by a {\em DUSTY} atmospheric model
(Baraffe et al.\ 1998), so we assume a model in which variability is
caused by clear ({\em COND}) ``holes'' in a photosphere dominated by
{\em DUSTY} ``clouds''. The {\em COND} spectrum is 0.55 mag brighter
than the {\em DUSTY} spectrum through our filter. The amplitude of
0.012\,mag we observe in Kelu-1 corresponds to {\em COND} ``holes''
covering $\sim$1.7\% of an otherwise {\em DUSTY} photosphere. This
value is a lower limit, as less perfect clear patches would have a
smaller effect on the variability, requiring them to cover larger
fractions of the surface.

Note that the covering fractions above actually represent the {\em
difference} in covering fraction between maximum and minimum
light. They are therefore only representative of surface features on
$\sim$180\deg~ scale. It may be that the surface has a far more
significant small scale structure, which simply averages out over
large regions.

\subsection{The radius of Kelu-1}
\label{sec:radius}

Interpreting the 1.8h variability as the rotation period allows us to make
one of the first firm tests of brown dwarf evolutionary
theory. Evolutionary models (e.g. Chabrier et al.\ 2000) predict that after
$\sim$100Myr, brown dwarfs (independant of mass) reach a stable radius of
$\sim$0.1R$_\odot$. We can combine the rotation period and the rotational
velocity to produce an observational lower limit (due to the unknown
inclination) on Kelu-1's radius. For the observed values of 1.80$\pm$0.04h
and 60$\pm$5kms$^{-1}$ (Basri et al.\ 2000), Kelu-1 must have $R \geq
0.09\pm0.01R_\odot$. The evolutionary models are therefore in agreement
with our observations of Kelu-1. This is not the case for the observations
of BRI0021-0214 made by Mart\'\i n et al.\ (2001). They require a radius of
at least 0.14R$_\odot$ to fit their observations with meta-stable
photospheric surface features. We note that this argument would require the
features responsible for photometric varibility to have the same rotation
period as the photosphere.

Alternatively, we can invert this argument and, assuming a model radius of
0.1$\pm$0.05R$_\odot$ (to include a range of possible ages and masses),
determine Kelu-1's inclination to the line of sight as $53 \leq i \leq
77$. This means that between 80\% and 97\% of the surface can cause
variability. The remaining fraction (polar regions) is either permanently
in view, or never visible.

\section{Conclusions}

It is now clear that variability from brown dwarfs can be detected, but
that photometry of better than 1\% is required to do so. In the near
future, futher studies of variability and rotation in brown dwarfs should
greatly increase our understanding of their physics.

We have detected a strong periodicity of 1.80$\pm$0.05 hours in
differential photometry of the L2 brown dwarf Kelu-1. We have investigated
four possible mechanisms to explain this variability:

\begin{enumerate}
\item surface inhomogeneity moderated by meteorology and variable dust
formation;
\item surface inhomogeneity moderated by magnetic starspots;
\item light curve variability due to gravitational distortion of Kelu-1's
envelope by a close companion; and
\item light curve variability due to an eclipsing binary
\end{enumerate}

Mechanisms (ii) \& (iv) seem unlikely explanations, but we are unable to
concusively differentiate between mechanisms (i) \& (iii). Ellipsodial
variability, mechanism (iii), would produce a twin peaked lightcurve,
giving a period of 3.6$\pm$0.1 hours. This mechanism will give stable and
repeatable photometric variability in future epochs. Alternatively,
mechanism (i) associates the 1.80$\pm$0.05 hour period with the rotation
period, which is consistent with the rotational velocity of 60 kms$^{-1}$
and theoretical radius of $\sim$0.1R$_\odot$, indicating an inclination in
the range 53\deg $\leq i \leq$ 77\deg.

Over the duration of our observations, the general shape and period of
the lightcurve are unchanged (at least to within the measurement noise),
implying the process causing the modulations is also stable on this
timescale. This is in contrast to previous variability observations where
no periodicity, or a period that changes on the order of the observation
length, has been measured.

The two explanations we have presented lead to different predictions for
future observations of Kelu-1. Variability induced by a secondary companion
will be completely repeatable at future epochs, whereas long term evolution
of surface features will result in a secular changes in the
lightcurve. Further observations of Kelu-1 will be a powerful discriminant
between the two hypotheses we have presented.

\section{Acknowledgements}

We would like to thank the referee France Allard for her speedy
reponse and very helpful comments. This paper is based on observations
made at Anglo-Australian Telescope, Siding Spring, Australia. FJC
acknowledges the support of a PPARC studentship award during the
course of this research.

\end{document}